\pdfoutput=1
\documentclass[11pt,twoside,a4paper,cmspaper,final,collab]{cms-tdr}
\def\svnVersion{0f73895}\def\svnDate{2022/06/01}\def\cmsCernNoTag{CERN-EP-2021-271}\def\cmsCernDate{\today}\def\cmsMessage{Published in the European Physical Journal C as \href{http://dx.doi.org/10.1140/epjc/s10052-022-10315-y}{\doi{10.1140/epjc/s10052-022-10315-y}.}}
\begin{document}\cmsNoteHeader{BPH-18-004}

\newcommand{\mupm}{\ensuremath{\PGmpm}\xspace}
\newcommand{\mumu}{\ensuremath{\MM}\xspace}
\newcommand{\pion}{\ensuremath{\PGp}\xspace}
\newcommand{\pip}{\ensuremath{\PGpp}\xspace}
\newcommand{\pim}{\ensuremath{\PGpm}\xspace}
\newcommand{\pipi}{\ensuremath{\pip\pim}\xspace}
\newcommand{\pipm}{\ensuremath{\PGppm}\xspace}
\newcommand{\kaon}{\ensuremath{\PK}\xspace}
\newcommand{\Kmp}{\ensuremath{\PKmp}\xspace}
\newcommand{\Kp}{\ensuremath{\PKp}\xspace}
\newcommand{\Km}{\ensuremath{\PKm}\xspace}
\newcommand{\KS}{\ensuremath{\PKzS}\xspace} 
\newcommand{\Kzero}{\ensuremath{\PKz}\xspace} 
\newcommand{\Kstar}{\ensuremath{\PKstP{892}^\pm}\xspace}
\newcommand{\Rho}{\ensuremath{\PGrP{770}^0}\xspace}
\newcommand{\jpsi}{\ensuremath{\PJGy}\xspace}
\newcommand{\psit}{\ensuremath{\PGyP{2\cmsSymbolFace{S}}}\xspace}
\newcommand{\Xm}{\HepParticleResonance{X}{3872}{}{}\Xspace}
\newcommand{\Xmm}{\HepParticleResonance{X}{4140}{}{}\Xspace}
\newcommand{\Ym}{\HepParticleResonance{Y}{4260}{}{}\Xspace}
\newcommand{\Zm}{\HepParticleResonance{Z}{4430}{}{+}\Xspace}
\newcommand{\B}{\ensuremath{\PB}\xspace}
\newcommand{\Bbar}{\ensuremath{\PAB}\xspace}
\newcommand{\Bd}{\ensuremath{\PBz}\xspace}
\newcommand{\Bs}{\ensuremath{\PBzs}\xspace}
\newcommand{\BsPsiKSKPi}{\ensuremath{\Bs\to\psit\KS\Kmp\pipm}\xspace}
\newcommand{\bran}{\ensuremath{\mathcal{B}}\xspace}
\newcommand{\ptb}{\ensuremath{\pt(\B)}\xspace}
\newcommand{\pvtx}{\ensuremath{P_{\text{vtx}}}\xspace}
\newcommand{\MKSPDG}{\ensuremath{m^{\text{PDG}}_{\Kzero}}\xspace}
\newcommand{\MPSITPDG}{\ensuremath{m^{\text{PDG}}_{\psit}}\xspace}
\newcommand{\dxy}{\ensuremath{D_{xy}}\xspace}
\newcommand{\dxyB}{\ensuremath{\dxy(\B)}\xspace}
\newcommand{\bdkpp}{\ensuremath{\Bd\to\psit\KS\pip\pim}\xspace}
\newcommand{\bdk}{\ensuremath{\Bd\to\psit\KS}\xspace}
\newcommand{\bsk}{\ensuremath{\Bs\to\psit\KS}\xspace}
\newcommand{\Nbsk}{\ensuremath{N(\bsk)}\xspace}
\newcommand{\Nbdk}{\ensuremath{N(\bdk)}\xspace}
\newcommand{\Nbdkpp}{\ensuremath{N(\bdkpp)}\xspace}
\newcommand{\splot}{\ensuremath{{}_{s}\mathcal{P}\text{lot}}\xspace}
\newcommand{\fsfd}{\ensuremath{f_\PQs/f_\PQd}\xspace}
\newcommand{\fdfs}{\ensuremath{f_\PQd/f_\PQs}\xspace}
\newcommand{\frfsfd}{\ensuremath{\frac{f_\PQs}{f_\PQd}}\xspace}
\newcommand{\frfdfs}{\ensuremath{\frac{f_\PQd}{f_\PQs}}\xspace}
\newcommand{\rs}{\ensuremath{R_{\PQs}}\xspace}
\newcommand{\rpp}{\ensuremath{R_{\pipi}}\xspace}

\ifthenelse{\boolean{cms@external}}{\providecommand{\cmsLeft}{upper\xspace}}{\providecommand{\cmsLeft}{left\xspace}}
\ifthenelse{\boolean{cms@external}}{\providecommand{\cmsRight}{lower\xspace}}{\providecommand{\cmsRight}{right\xspace}}

\cmsNoteHeader{BPH-18-004}

\title{Observation of \texorpdfstring{\bdkpp}{B0 to psi(2S) K0S pi+pi-} and \texorpdfstring{\bsk}{B0s to psi(2S) K0S} decays}

\date{\today}

\abstract{
\tolerance=1800
Using a data sample of $\sqrt{s}=13\TeV$ proton-proton collisions  collected by the CMS experiment at the LHC in 2017 and 2018 with  an integrated luminosity of 103\fbinv, the \bsk and \bdkpp decays  are observed with significances exceeding 5 standard deviations. The resulting branching fraction ratios, measured for the first time, correspond to $\bran(\bsk)/\bran(\bdk) = (3.33 \pm 0.69 \stat \pm 0.11\syst \pm 0.34\,(\fsfd)) \times 10^{-2}$ and  $\bran(\bdkpp)/\bran(\bdk) = 0.480 \pm 0.013\stat \pm 0.032\syst$, where the last uncertainty in the first ratio is related to the uncertainty in the ratio of production  cross sections of \Bs and \Bd mesons, \fsfd.
\par
}

\hypersetup{%
pdfauthor={CMS Collaboration},%
pdftitle={Observation of B0 to psi(2S) K0S pi+ pi- and B0s to psi(2S) K0S decays},%
pdfsubject={CMS},%
pdfkeywords={CMS, B mesons, spectroscopy, exotic hadrons, experimental results}}

\maketitle

\section{Introduction}

{\tolerance=400 
Decays of neutral \B mesons into charmonium resonances (\jpsi, \psit, etc.) are 
well suited to study the flavour sector of the standard model
(SM) and to search for indications of new physics beyond the SM. 
In the last decade, interest in \PQb hadron decays to final states containing a 
charmonium resonance has increased after several exotic hadrons have been observed
as intermediate resonances in multibody decays. 
Starting from the observation of \Xm~\cite{Choi_2003}, 
many new charmonium-like states have been observed, such as \Xmm~\cite{CDF:2009jgo,CDF:2011pep,CMS:2013jru,LHCb:2016nsl}, 
\Ym~\cite{BaBar:2005hhc,BESIII:2016bnd},
and others, with properties (mass, width and decay pattern) not 
fitting into the landscape of traditional charmonium states. 
The first charged tetraquark candidate, \Zm, was discovered in the $\B\to\psit\kaon\pip$ 
decay as a peak in the $\psit\pip$ mass spectrum~\cite{Belle:2007hrb,Belle:2009lvn,Belle:2013shl,LHCb:2014zfx}.
Many other exotic hadrons have been observed in the last 15 years~\cite{Hosaka:2016pey,PDG},
and the nature of most of them is still unclear. 
Moreover, channels whose final state is accessible both from \B and \Bbar can be
used to measure time-dependent \textit{CP} 
asymmetry~\cite{PhysRevLett.10.531, Belle:2001zzw, BaBar:2001pki, BaBar:2008kfx, BaBar:2009byl, Belle:2012paq, LHCb:2014xpr, LHCb:2015ups, LHCb:2015brj, LHCb:2017mpa, Belle:2018nxw, LHCb:2019nin, CMS:2020efq, ATLAS:2020lbz} as well.
\par}

This paper presents the first measurement of the \bsk and \bdkpp
decays, using a data sample of proton-proton collisions at $\sqrt{s}=13\TeV$ collected by the CMS experiment at 
the CERN LHC in 2017 and 2018 with an integrated luminosity of 103\fbinv~\cite{CMS:2018elu,CMS:2019jhq}. 
Both decays can potentially be used for \textit{CP} asymmetry measurements, 
and, in addition, the second one can also be used to search for intermediate exotic resonances.
The \psit and \KS mesons are reconstructed using their decays into \mumu 
and \pipi, respectively. 
The \bdk decay is chosen as the normalization channel for the measurement of the 
branching fractions, since  
its probability is precisely known~\cite{PDG}, and its topology and kinematic properties 
are similar to those of the \bsk or \bdkpp decays. Therefore, using this
normalization reduces the systematic uncertainties related to muon and track reconstruction. 
The relative branching fractions are measured using the relations 
\begin{linenomath}
\ifthenelse{\boolean{cms@external}}
{ % Begin journal version
    \begin{equation} 
    \label{eq:eqbr}
    \begin{aligned}
       \rs  \equiv& \frac{\bran(\bsk)}{\bran(\bdk)} \\ 
             =& \frfdfs \, \frac{\epsilon(\bdk)}{\epsilon(\bsk)} \, \frac{\Nbsk}{\Nbdk},           \\
      \rpp  \equiv& \frac{\bran(\bdkpp)}{\bran(\bdk)}  \\
             =& \frac{\epsilon(\bdk)}{\epsilon(\bdkpp)}\\
            &\times \frac{\Nbdkpp}{\Nbdk},
    \end{aligned}
    \end{equation}
} % End journal version
{ % Begin arXiv version
    \begin{equation} 
    \label{eq:eqbr}
    \begin{aligned}
        \rs &\equiv \frac{\bran(\bsk)}{\bran(\bdk)} = \frfdfs \, \frac{\epsilon(\bdk)}{\epsilon(\bsk)} \, \frac{\Nbsk}{\Nbdk},           \\
        \rpp &\equiv \frac{\bran(\bdkpp)}{\bran(\bdk)} = \frac{\epsilon(\bdk)}{\epsilon(\bdkpp)} \, \frac{\Nbdkpp}{\Nbdk},
    \end{aligned}
    \end{equation}
} % End arXiv version
\end{linenomath}
where \bran is the branching fraction, $N$ is the number of reconstructed events in data, $\epsilon$ is 
the total reconstruction efficiency, and \fdfs is the ratio of production 
cross sections of \Bd and \Bs mesons (also called 
fragmentation fraction ratio).
Charge-conjugate states are implied to be included throughout the paper.

Tabulated results are provided in the HEPData record for this analysis~\cite{hepdata}.

\section{The CMS detector and simulated event samples\label{sec2}}

The central feature of the CMS apparatus~\cite{Chatrchyan:2008zzk} 
is a superconducting solenoid of 
6\unit{m} internal diameter, providing a magnetic field of 3.8\unit{T}.
Within the solenoid volume are a silicon pixel and strip tracker, a lead 
tungstate crystal electromagnetic calorimeter, and a brass and scintillator 
hadron calorimeter. Muons are measured in gas-ionization detectors embedded in 
the steel flux-return yoke outside the solenoid. 

Events of interest are selected using a two-tiered trigger system~\cite{Khachatryan:2016bia}. 
The first level, composed of custom hardware processors, uses information from the 
calorimeters and muon detectors to select events at a rate of around 100\unit{kHz} within a time 
interval of less than 4\mus~\cite{Sirunyan:2020zal}. 
The first-level trigger used in this analysis requires at least two muons. 
The second level, known as the high-level trigger, consists of a farm of processors running 
a version of the full event reconstruction software optimized for fast processing that reduces the 
event rate to around 1\unit{kHz} before data storage. The high-level trigger algorithm used in the analysis 
requires two opposite-sign muons compatible with the dimuon decay of a \psit meson 
with transverse momentum (\pt) larger than 18\gev.

Simulated Monte Carlo samples for the decays of interest are generated for the analysis. 
The \PYTHIA 8.230 package~\cite{Pythia8} with the CP5 tune~\cite{CMS:2019csb} is used 
to simulate the production of the \Bd and \Bs mesons,
whose subsequent decays are performed by \EVTGEN 1.6.0~\cite{Lange2001152}, where final-state photon 
radiation is included using \PHOTOS 3.61~\cite{BARBERIO1991115,BARBERIO1994291}.
The lifetimes of \Bd and \Bs mesons used in the generation are 1.52 and 1.47\unit{ps}, respectively. 
The generated events are passed to a detailed \GEANTfour-based
simulation~\cite{Agostinelli2003250} of the CMS detector, and are then processed using 
the same trigger and reconstruction as used for the collision data.
The simulation includes effects from multiple proton-proton interactions in the same
or nearby bunch crossings (pileup) with the multiplicity distribution tuned to match those of the data.

\section{Event reconstruction and selection\label{sec3}}

{\tolerance=2000 
The reconstruction procedure starts with finding two muons of opposite charges, 
that must match those that triggered the event readout. 
The muon candidates are required to have $\pt(\mupm)>3\gev$, a pseudorapidity 
$\abs{\eta(\mupm)}<2.4$, and to satisfy general identification (soft-muon) criteria~\cite{CMS:2018rym}. 
The two muons with a two-prong vertex fit probability $\pvtx(\mumu) > 1\%$ are 
paired to form the \psit candidate, which must have $\pt(\psit)>18\gev$ and an 
invariant mass $3500<m(\mumu)<3950\MeV$ (the world average \psit meson mass is 
$\MPSITPDG = 3686.10\MeV$~\cite{PDG}).
\par}

{\tolerance=800 
The $\KS\to\pipi$ candidates are formed from displaced two-prong vertices,
as described in Ref.~\cite{Khachatryan:2010pw}.
The \pipi invariant mass is required to be within $\pm20\MeV$ 
of the world average value $\MKSPDG=497.611\MeV$~\cite{PDG}, which corresponds to
approximately three times the mass resolution. Selected 
\pip and \pim tracks are then refitted with their invariant mass constrained to \MKSPDG, and the obtained 
\KS candidate is required to have $\pvtx(\pipi) > 1\%$.
\par}

{\tolerance=800 
The $\B\to\psit\KS$ candidates are obtained through a kinematic vertex fit 
on the $\mumu\KS$ system which constrains the dimuon mass to \MPSITPDG.
The \KS candidates are required to have $\pt(\KS)>1\gev$, a 3D pointing 
angle between $\vec{D}(\KS)$ and $\vec{p}(\KS)$ to satisfy 
$\cos(\vec{D}(\KS),\,\vec{p}(\KS))>0.99$, and a transverse 
displacement significance for \KS of $\dxy > 5 \sigma_{\dxy}$.
Here $\vec{D}(\KS)$ denotes the vector from the \KS production vertex to the \KS decay 
vertex, while \dxy and $\sigma_{\dxy}$ correspond to the length 
of $\vec{D}_{xy}$, the transverse component of $\vec{D}$, and its uncertainty. 
To suppress the combinatorial background, additional requirements are applied:
$\pvtx(\mumu \KS)>5\%$, $\cos(\vec{D}_{xy}(\B),\,\vec{\pt}(\B))>0.99$, 
and $\dxyB > 5 \sigma_{\dxyB}$, where the \B meson transverse displacement \dxyB is calculated 
with respect to the primary vertex (PV).
From all reconstructed proton-proton collision points, the PV is chosen as the one 
with the smallest \B pointing angle, as in Refs.~\cite{CMS:2017hfy,CMS:2019kbn,CMS:2019mny}. 
The pointing angle is the angle formed by the \B candidate momentum and 
the vector from the PV to the reconstructed \B candidate vertex.
Furthermore, if in this procedure any of the tracks used in the 
\B candidate reconstruction is included in the fit of the chosen 
PV, the track is removed, and the PV is refitted.
\par}

For the \bdkpp candidates, two additional, oppositely charged, high-purity~\cite{Chatrchyan:2014fea}
tracks, assumed to be pions and having $\pt>0.9\gev$, are included in the \B meson vertex fit, 
while the rest of the selection criteria are the same. 

\section{Observation of the \texorpdfstring{\bsk}{B0s to psi(2S) K0S} decay
\label{sec4}}

The measured $\psit\KS$ invariant mass distribution is presented in Fig.~\ref{fig:Signals} (\cmsLeft).
The \Bd signal (left peak) is described with a double Gaussian function with common mean, 
whose parameters are free to vary in an unbinned maximum-likelihood fit. 
It is found in simulation that the \bsk signal (right peak) has the same 
shape as the \bdk signal, but it is about 10\% wider, because of the larger energy release in the decay.
Therefore, the \Bs signal is modelled with a double Gaussian 
function of the same shape as the \Bd signal, with the resolution parameters 
scaled by the ratio of the widths found in the simulation. 
The background is modelled with an exponential function.
The good quality of the fit is verified by calculating the $\chi^2$ 
between the binned distribution and the fit function, resulting in 
$\chi^2=83$ for 91 degrees of freedom.

The ratio of signal yields 
\ifthenelse{\boolean{cms@external}}
{ % Begin journal version
$\Nbsk/$ \\ $\Nbdk = (6.8 \pm 1.4)\times 10^{-3}$ 
} % End journal version
{ % Begin arXiv version
$\Nbsk/\Nbdk = (6.8 \pm 1.4)\times 10^{-3}$ 
} % End arXiv version
is extracted from the fit. 
Its uncertainty is calculated by taking into account the correlation between the uncertainties in \Bs and \Bd yields, 
which are found to be $113 \pm 23$ and $16660 \pm 140$, respectively, where the uncertainties are statistical only.

{\tolerance=1600 
The statistical significance of the \bsk signal is evaluated with the likelihood ratio technique, comparing the background-only and signal-plus-background hypotheses, 
with the standard asymptotic formula~\cite{Cowan:2010js}, assuming that the conditions to apply Wilks' theorem~\cite{Wilks:1938dza} are satisfied.
For a significance estimation, the mass difference between the \Bs and \Bd signals is fixed to the known value of 83.78\MeV~\cite{PDG}. 
The obtained significance is 5.2 standard deviations and varies in the 
range 5.1--5.4 standard deviations when accounting for the systematic 
uncertainties due to the choice of the fit model, discussed in Section~\ref{sec:syst}.
\par}

\section{Observation of the \texorpdfstring{\bdkpp}{B0 to psi(2S) K0S pi+ pi-} decay
\label{sec5}}

\begin{figure}[h]
    \centering
    \includegraphics[width=0.49\textwidth]{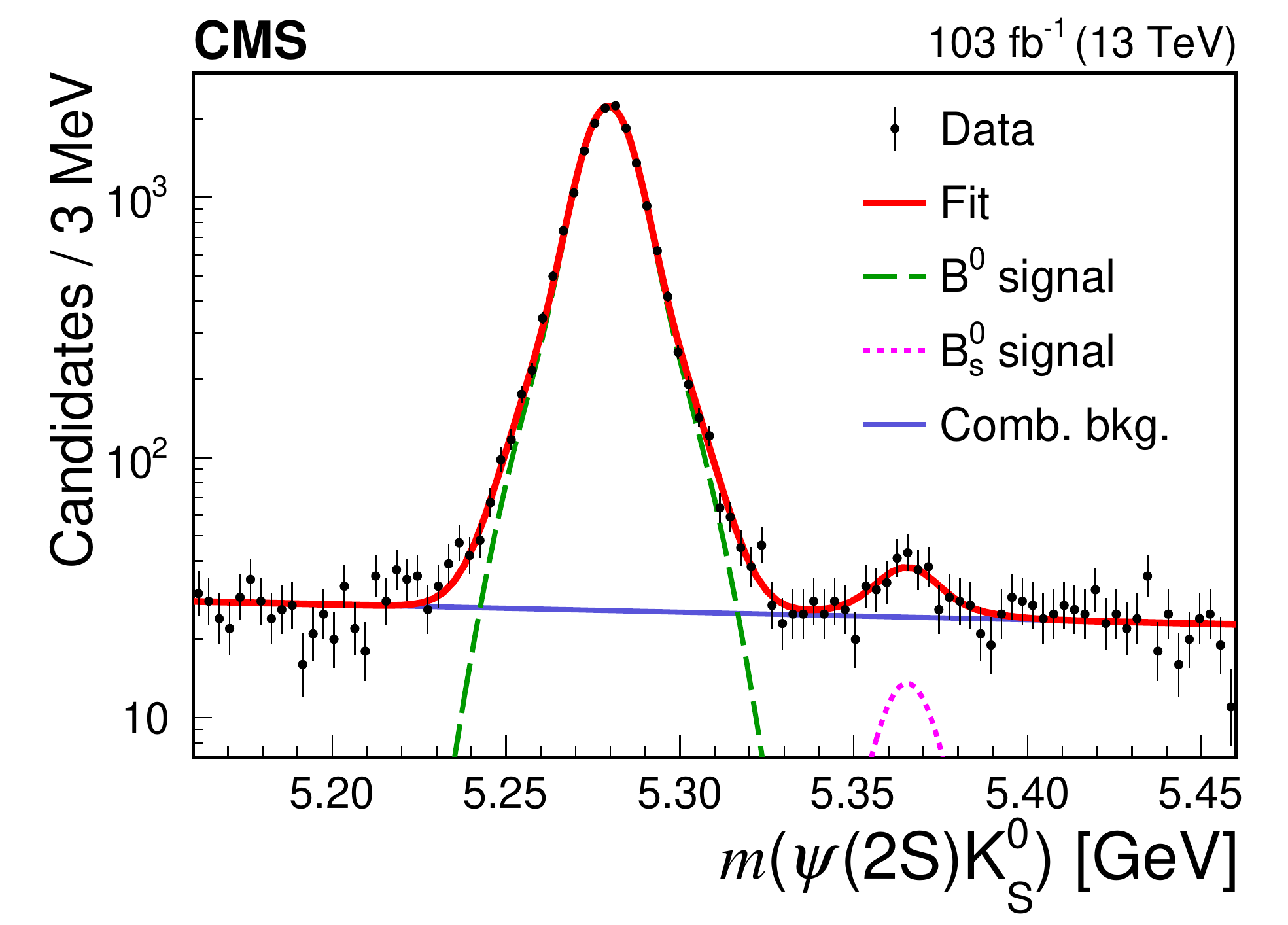}
    \includegraphics[width=0.49\textwidth]{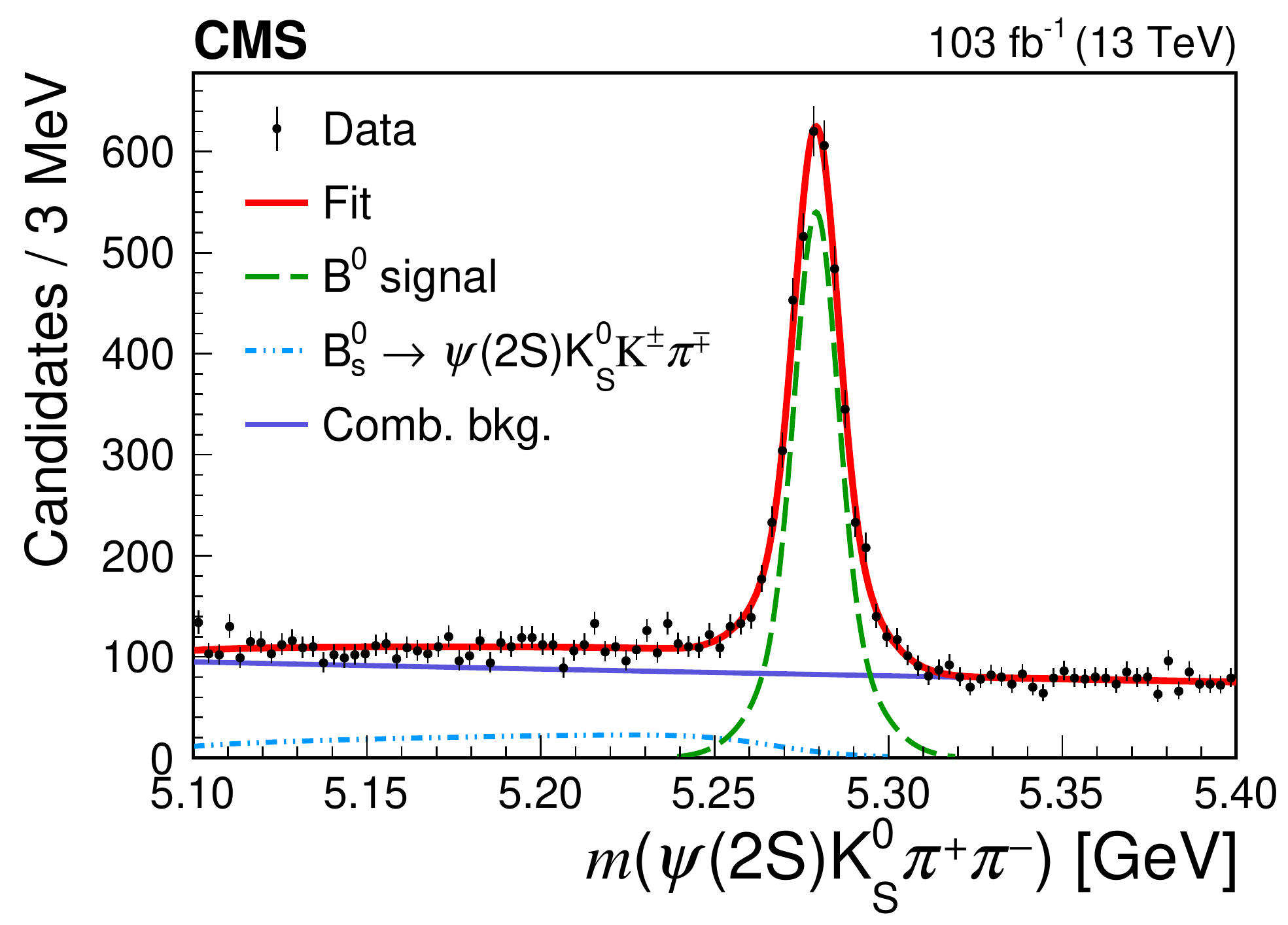}

    \caption{Measured invariant mass distributions of $\psit\KS$ (\cmsLeft) and $\psit\KS\pipi$ (\cmsRight) candidates.
    The overlaid results from the fit are described in the text.}
    \label{fig:Signals}
\end{figure}

As shown in Fig.~\ref{fig:Signals} (\cmsRight), the measured $\psit\KS\pipi$ mass distribution
presents a clear \bdkpp signal peak on top of a relatively small background. 
The \Bd signal is modelled with a double Gaussian function with common mean with all parameters free to vary,
and the combinatorial background is described by an exponential function.

Studies of simulated events show that the \BsPsiKSKPi decay contributes to the reconstructed 
$\psit\KS\pipi$ mass distribution when the charged kaon is reconstructed as a pion. 
This relevant background contribution is accounted for in the fit to data by including a 
dedicated component with a freely varying normalization and a fixed shape that is obtained from simulation (Fig.~\ref{fig:Signals}, \cmsRight). 

The signal yield \Nbdkpp is found to be $3498 \pm 87$, where the uncertainty is statistical only.
The $\chi^2$ between the binned distribution and 
the fit function is 75 for 92 degrees of freedom, demonstrating the good quality of the fit.
The significance of the \bdkpp signal, evaluated as described in Section~\ref{sec4}, exceeds 30 
standard deviations.

\begin{figure*}[h]
\centering
    \includegraphics[width=0.43\textwidth]{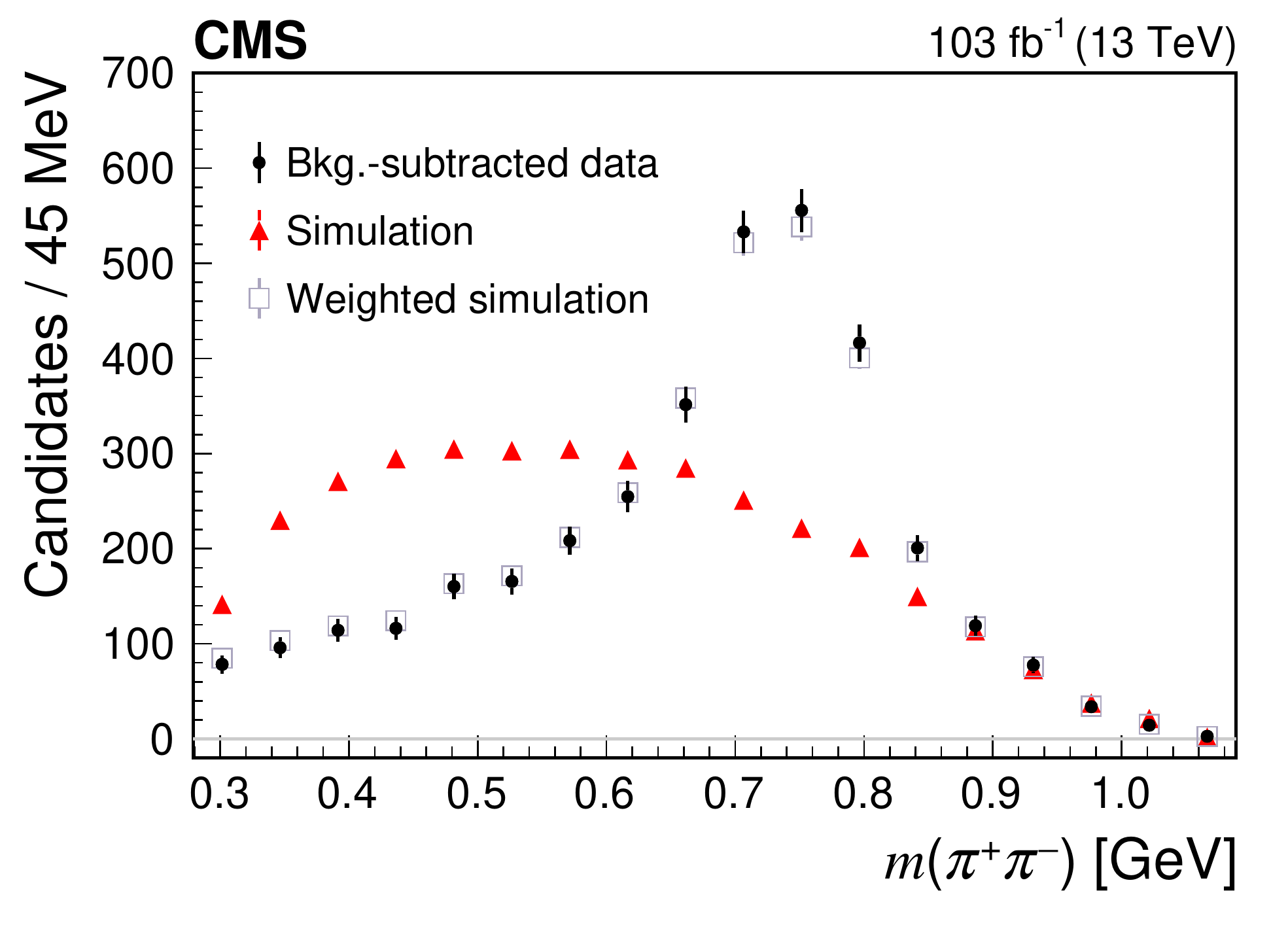}
    \includegraphics[width=0.43\textwidth]{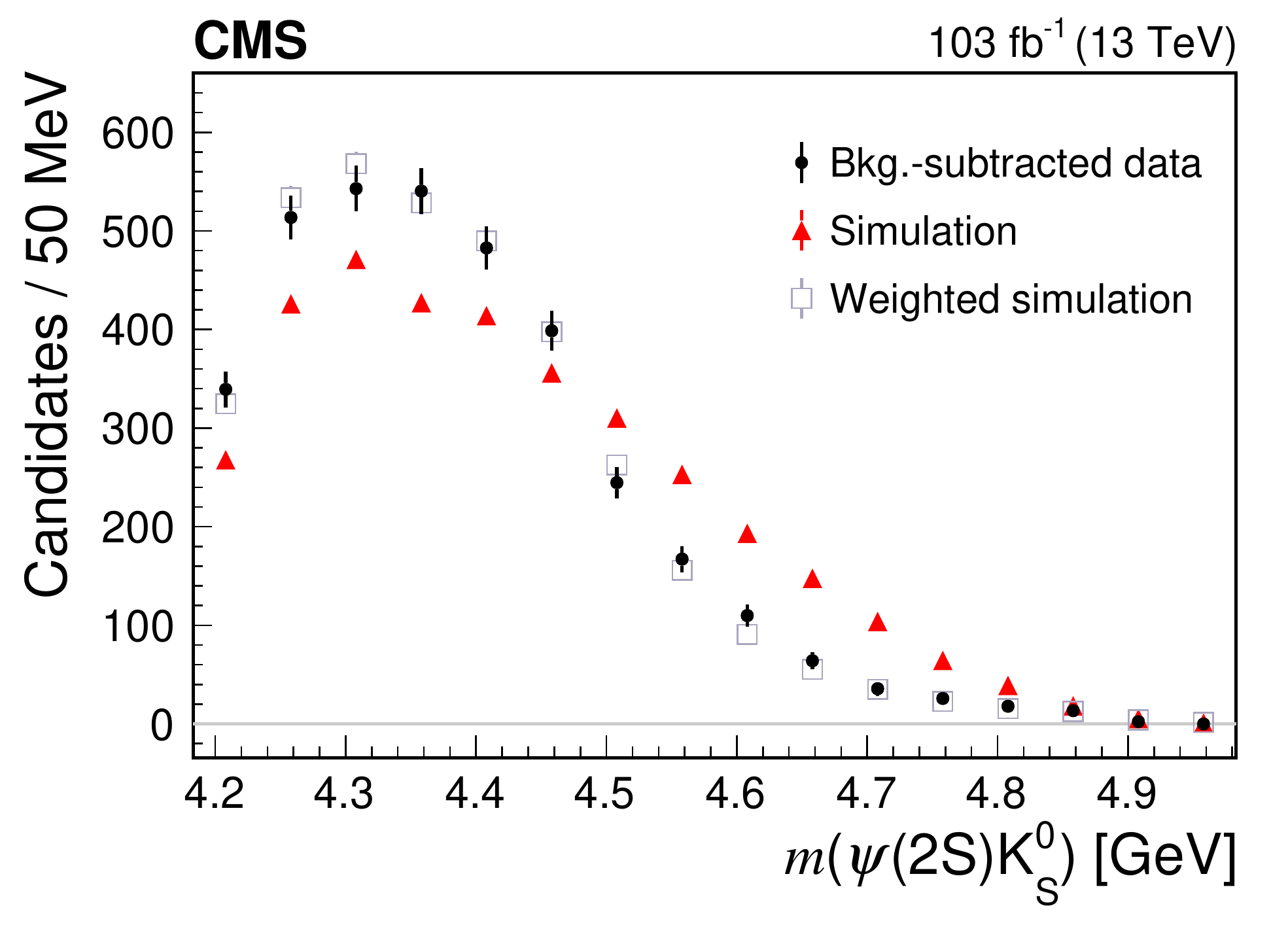}
    \includegraphics[width=0.43\textwidth]{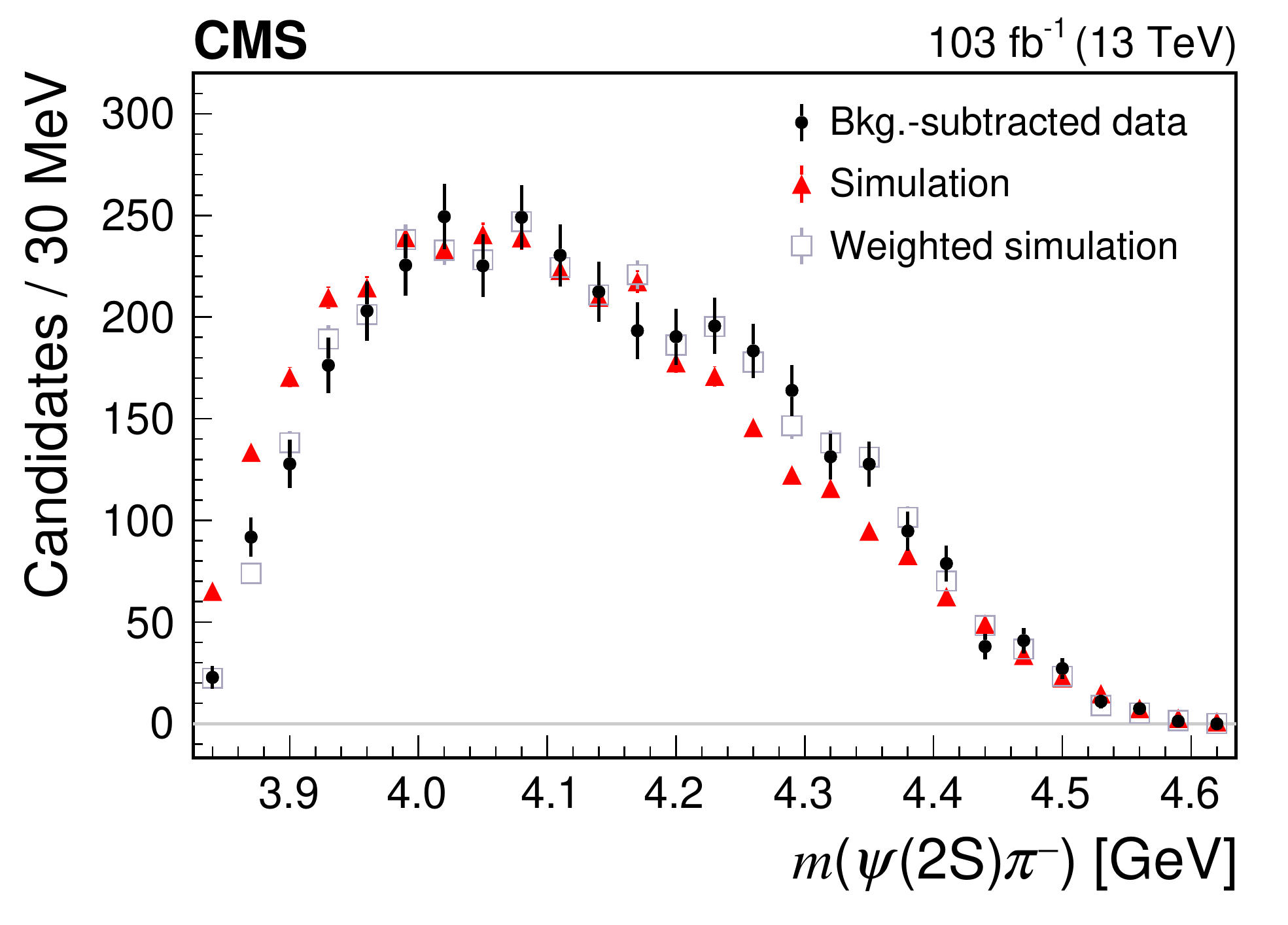}
    \includegraphics[width=0.43\textwidth]{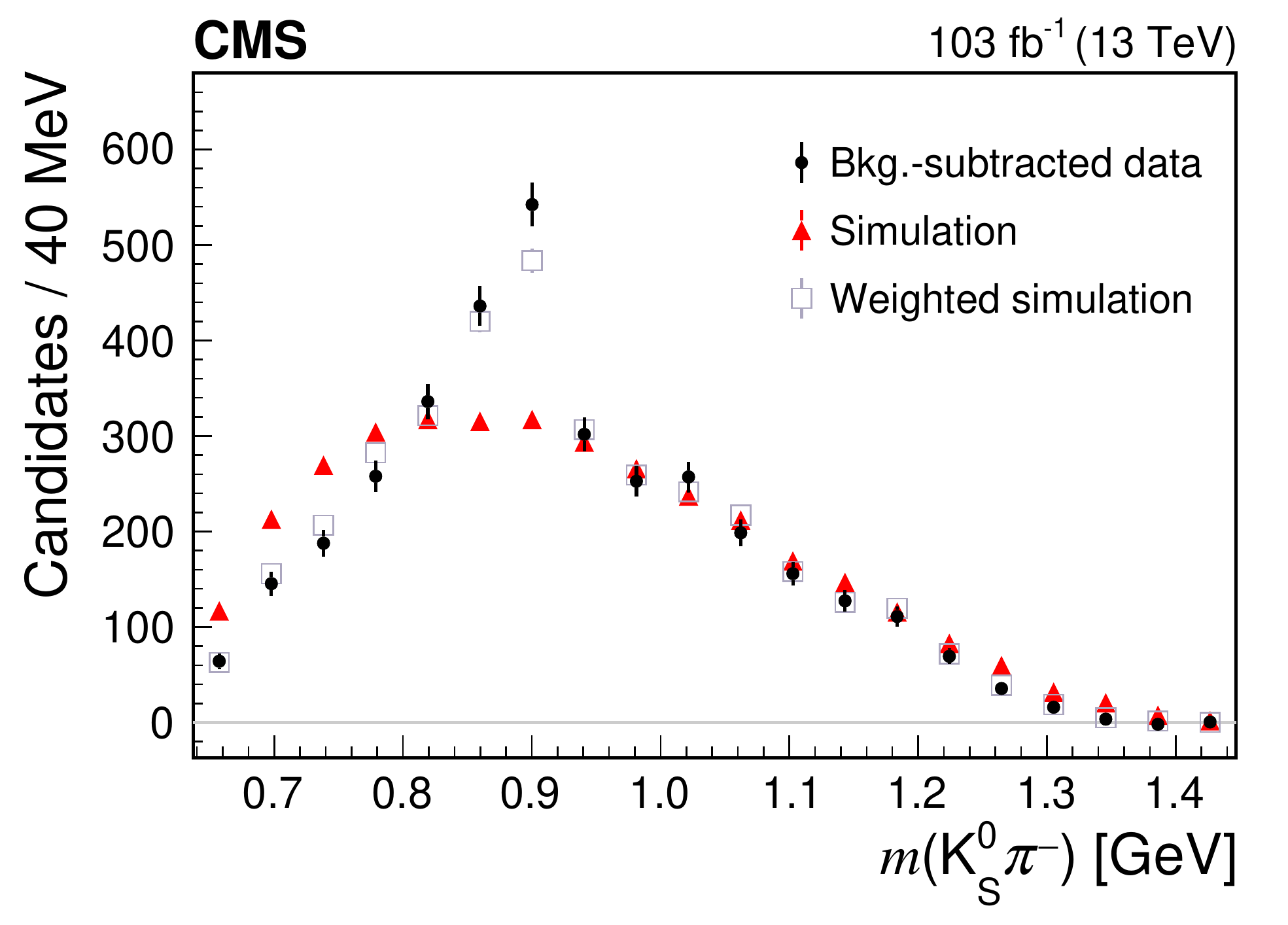}
    \includegraphics[width=0.43\textwidth]{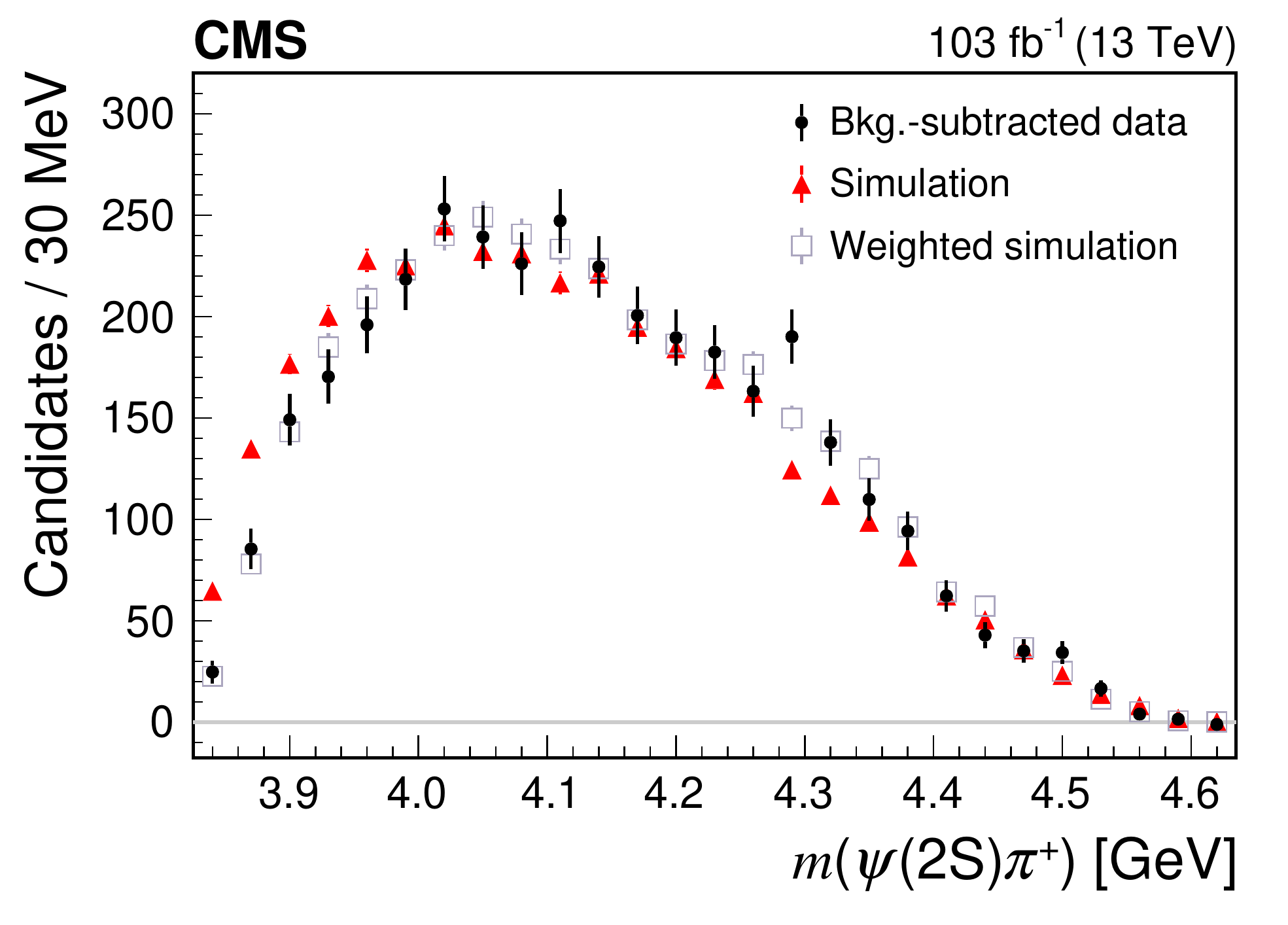}
    \includegraphics[width=0.43\textwidth]{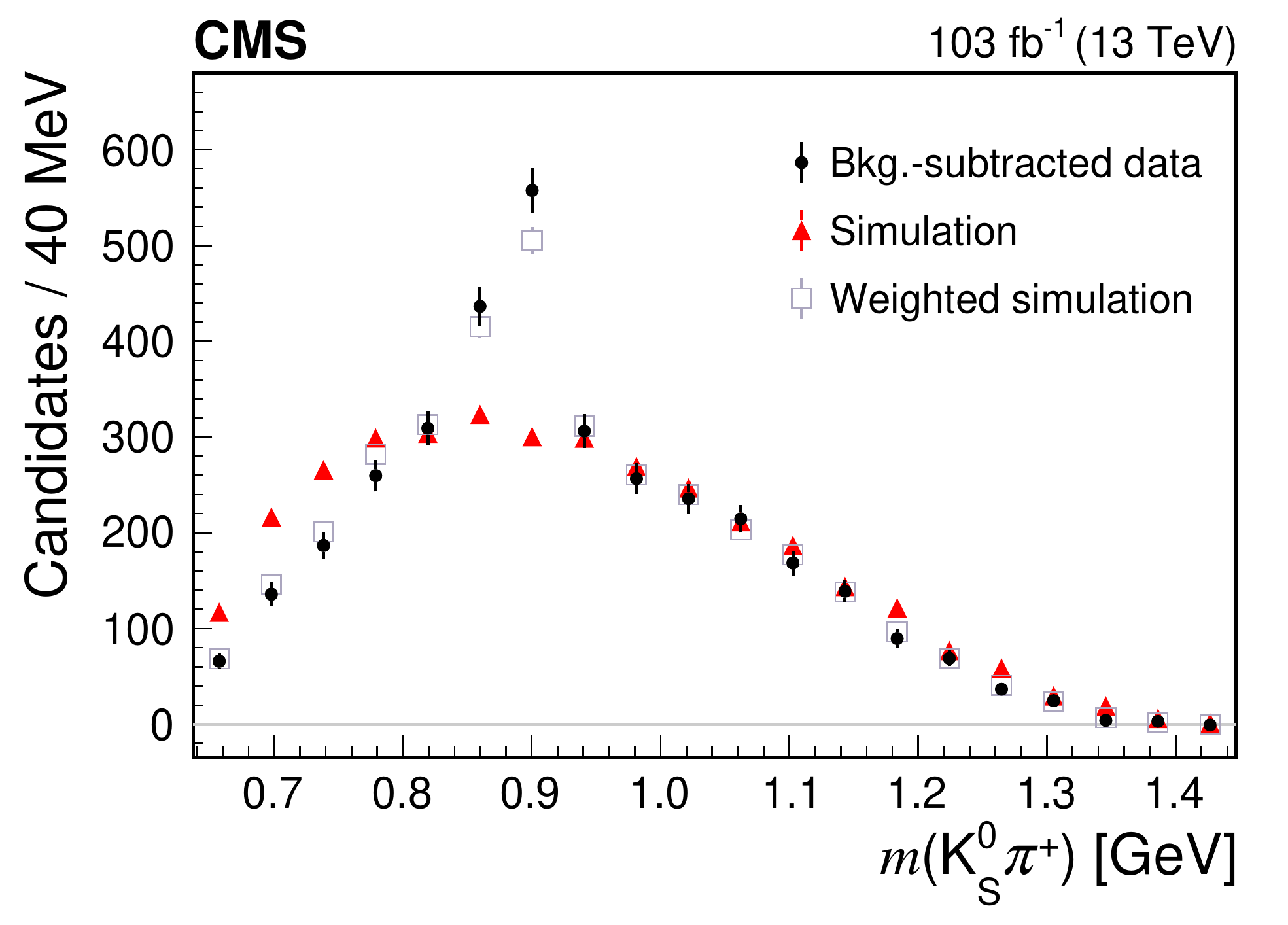}
    \caption{Distributions of 2-body intermediate invariant masses from the \bdkpp decay. 
    The data distributions (black dots) are background subtracted.
    Overlaid are the predictions of phase space simulations (red triangles), as well as the predictions after 
    applying the reweighting procedure described in Section~\ref{sec:syst} (grey squares). 
    }
    \label{fig:Rewhists2}
\end{figure*}

\begin{figure*}[h!t]
\centering
    \includegraphics[width=0.43\textwidth]{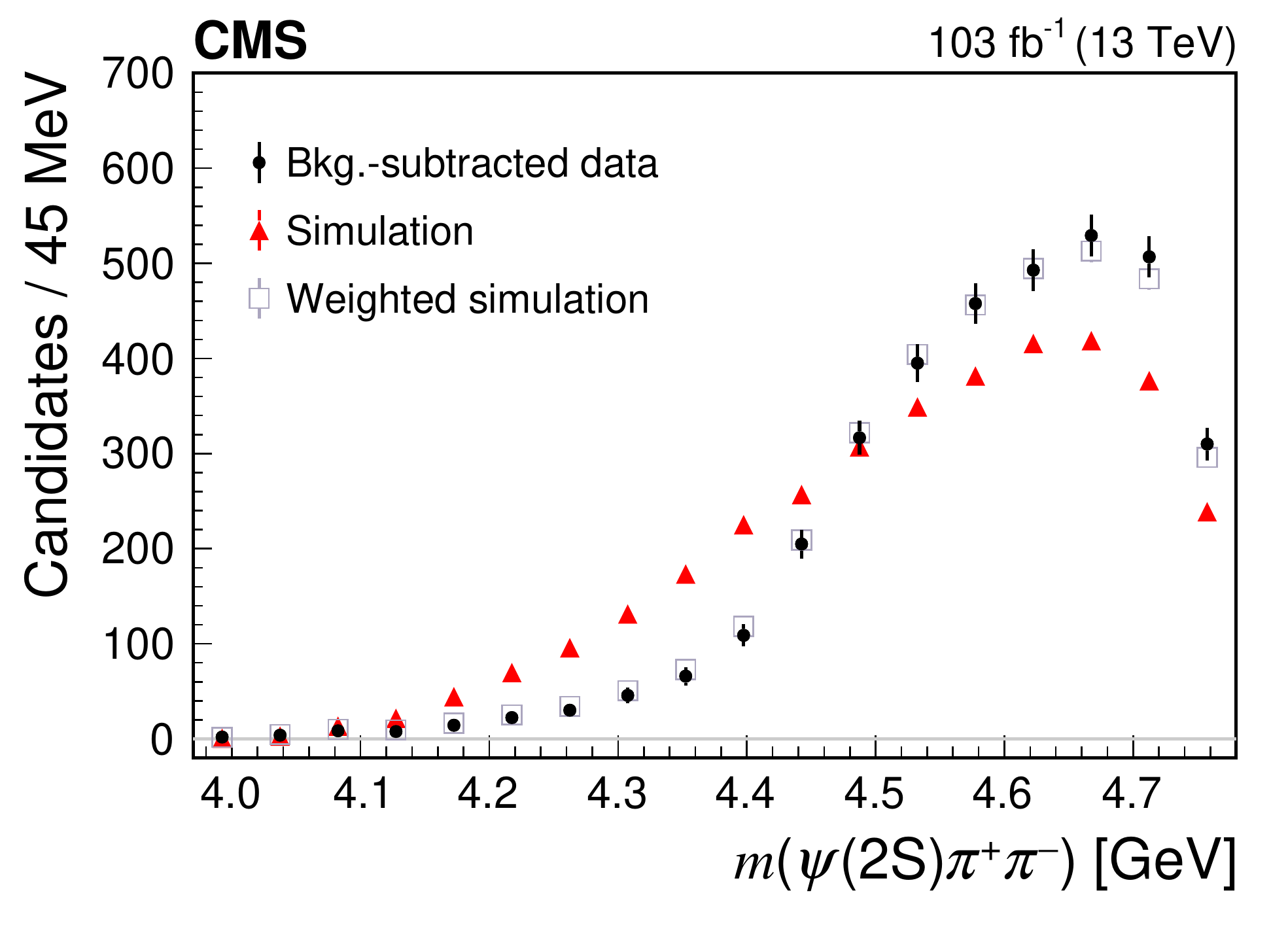}
    \includegraphics[width=0.43\textwidth]{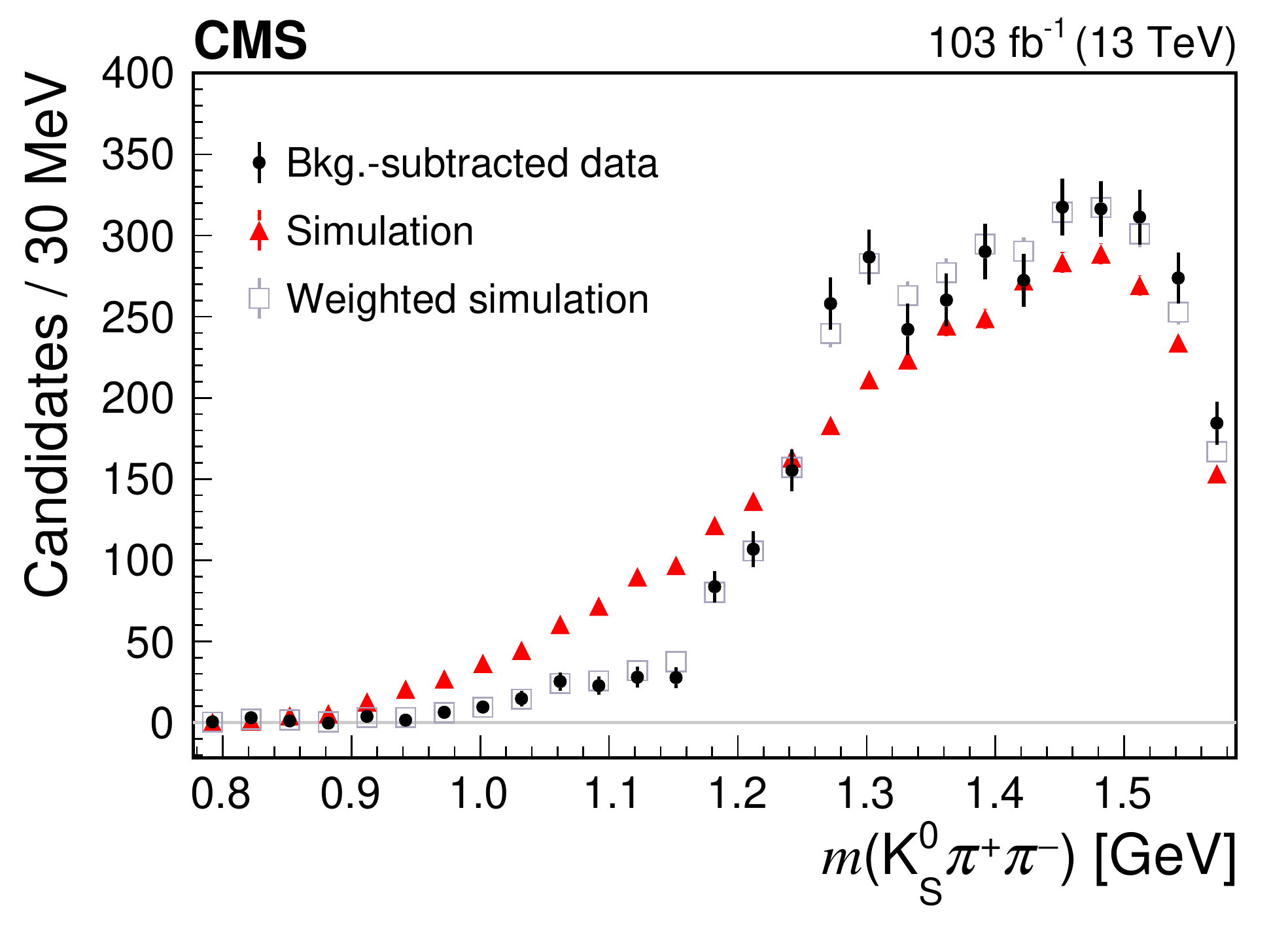}
    \includegraphics[width=0.43\textwidth]{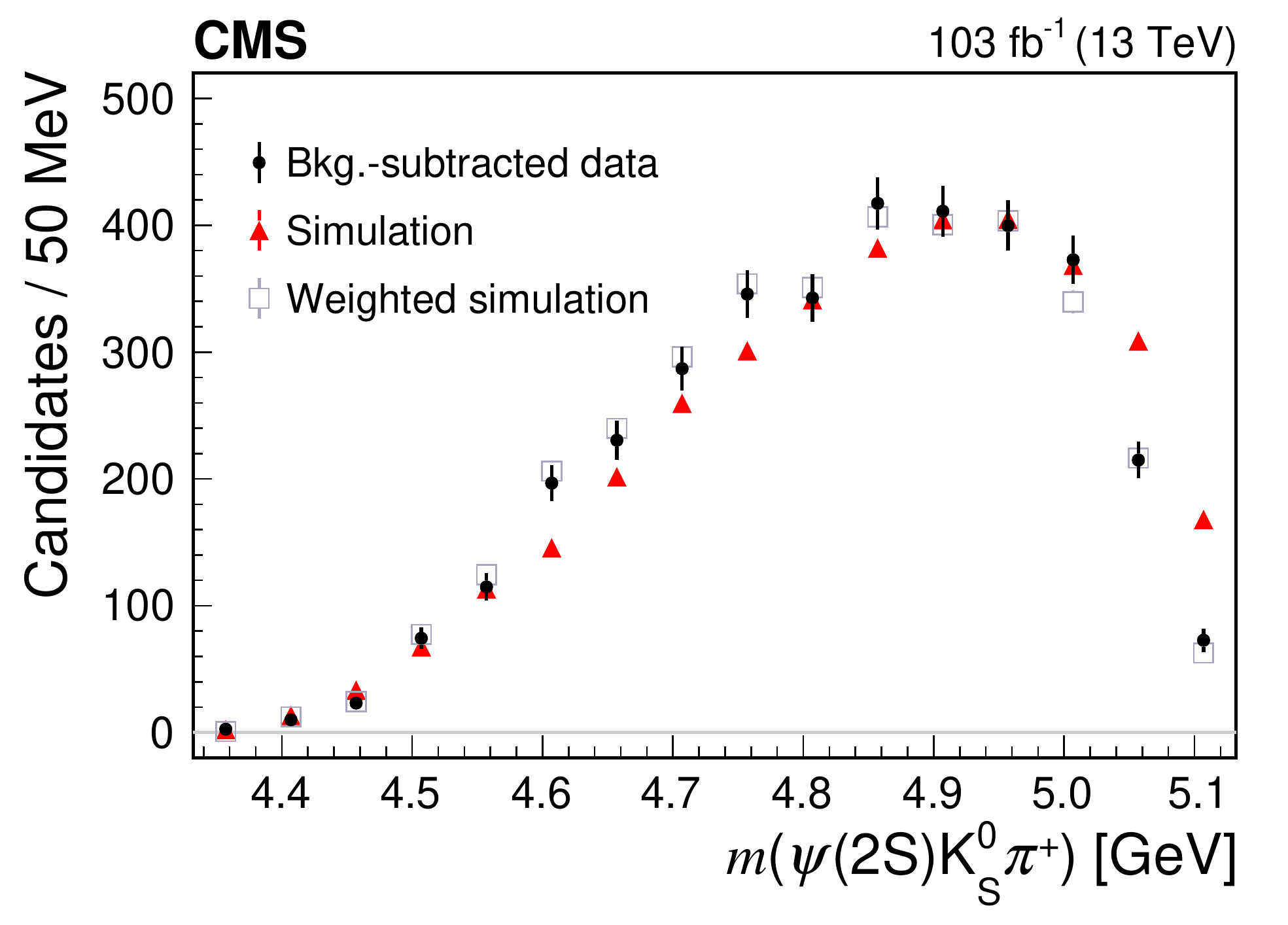}
    \includegraphics[width=0.43\textwidth]{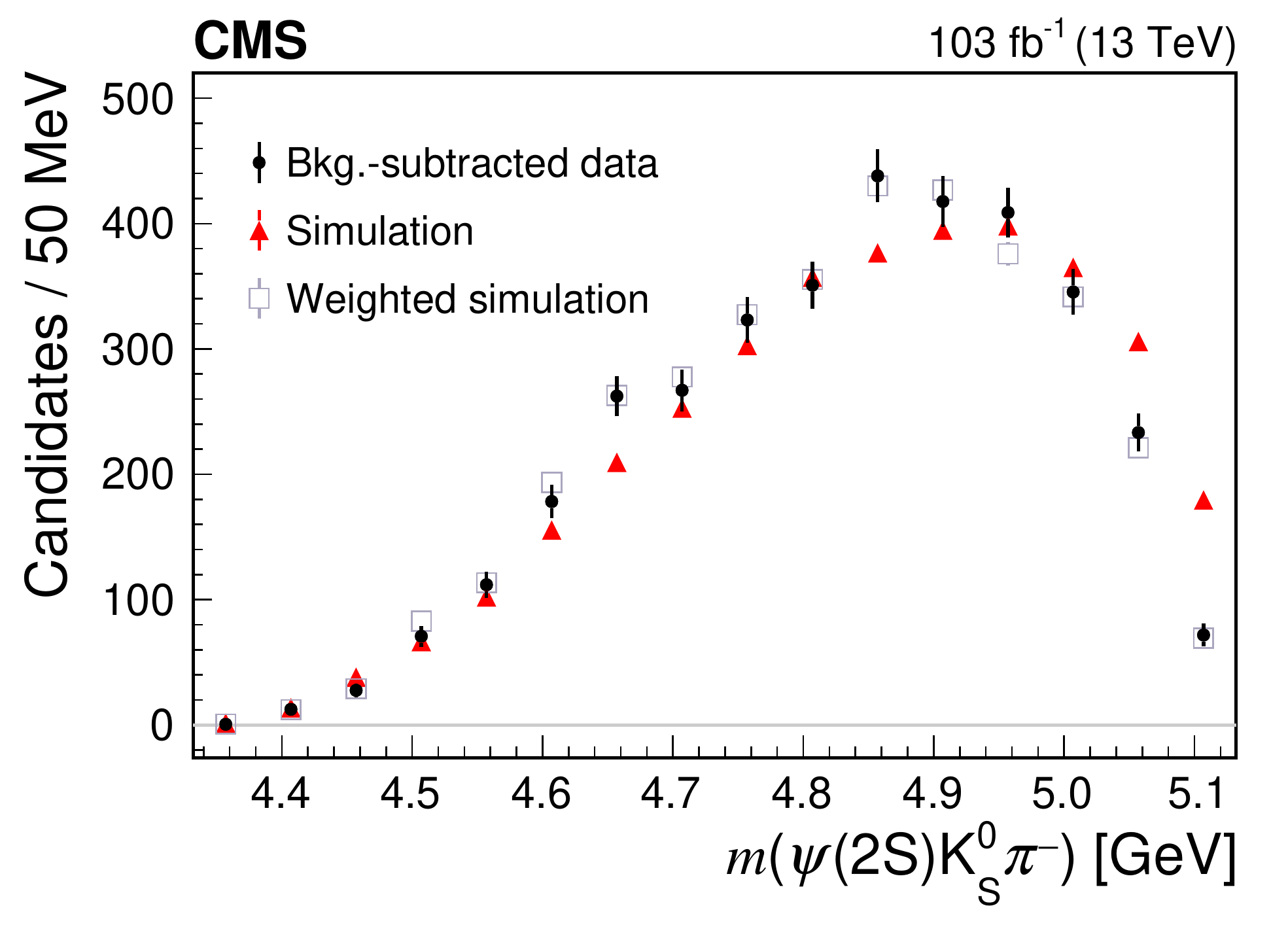}
    \caption{Distributions of 3-body intermediate invariant masses from the \bdkpp decay. 
    Data distributions (black dots) are background subtracted.
    Overlaid are the predictions of phase space simulations (red triangles), as well as the predictions after 
    applying the reweighting procedure described in Section~\ref{sec:syst} (grey squares). 
    }
    \label{fig:Rewhists3}
\end{figure*}

{\tolerance=5500 
The intermediate invariant mass distributions, corresponding to 
the four-body \bdkpp decay, 
are produced using the \splot~\cite{Pivk:2004ty} technique to subtract the non-\Bd background, 
using the $m(\psit\KS\pipi)$ distribution fit described above. The correlations 
between the intermediate invariant masses and $m(\psit\KS\pipi)$ have been checked to be below 10\%. 
Figures~\ref{fig:Rewhists2} and~\ref{fig:Rewhists3} show the 2- and 3-body invariant mass distributions. 
Overlaid are the predictions of the 4-body phase space simulations, which provide poor description of the data since the simulations do not account for the intermediate resonance structure.
The simulation after application of the reweighting
procedure described in Section~\ref{sec:syst} is also shown. 
The mass distributions of \psit and one or two light mesons 
($\psit\KS$, $\psit\pipm$, $\psit\KS\pipm$, $\psit\pipi$) 
do not present any significant 
narrow peak that could indicate a contribution from an exotic charmonium state.
The small excess at about 4.3\GeV in the $m(\psit\pip)$ distribution (Fig.~\ref{fig:Rewhists2}, bottom left) 
is not significant, and there is no similar excess in the $m(\psit\pim)$ distribution 
(Fig.~\ref{fig:Rewhists2}, middle left).
Moreover, exotic states previously found in this mass range are known to have 
large natural widths~\cite{Hosaka:2016pey,PDG}. 
Signs of the \Kstar (Fig.~\ref{fig:Rewhists2}, middle and bottom right), 
\Rho (Fig.~\ref{fig:Rewhists2}, top left), 
and $\PKDoP{1270}^0$ (Fig.~\ref{fig:Rewhists3}, top right) resonances are seen in the 
mass distributions of $\KS\pipm$, \pipi, and $\KS\pipi$, respectively.
\par}

\section{Efficiencies}
\label{sec:eff}

The total reconstruction efficiency for each decay channel is evaluated using samples of simulated events.
It is calculated as the number of reconstructed events divided by the number of 
generated events, and includes the detector acceptance, trigger, 
and candidate reconstruction efficiencies.
Only the ratios of such efficiencies are needed to measure the ratios \rs and \rpp, 
thus reducing the systematic uncertainties associated with muon and track
reconstruction. 

The obtained efficiency ratios are found to be
\begin{linenomath}
\ifthenelse{\boolean{cms@external}}
{ % Begin journal version
    \begin{equation*}
    \begin{aligned}
        \label{eqbr}
        \frac{\epsilon(\bdk)}{\epsilon(\bsk)} &= 1.019 \pm 0.013\text{~and}\\
        \frac{\epsilon(\bdk)}{\epsilon(\bdkpp)} &= 2.288 \pm 0.026,
    \end{aligned}
    \end{equation*}
} % End journal version
{ % Begin arXiv version
    \begin{equation*}
        \label{eqbr}
        \frac{\epsilon(\bdk)}{\epsilon(\bsk)} = 1.019 \pm 0.013\text{~~and~~}\frac{\epsilon(\bdk)}{\epsilon(\bdkpp)} = 2.288 \pm 0.026,
    \end{equation*}
} % End arXiv version
\end{linenomath}
where the uncertainties are statistical only and are related to the size of
the simulated event samples. The first ratio is close to unity, as expected, 
while the second ratio is significantly greater than unity because of the presence 
of two additional tracks in the denominator. 
The lifetimes of heavy and light \Bs meson eigenstates differ by about 0.2\unit{ps}~\cite{PDG}, 
which can have an impact on the efficiency $\epsilon(\bsk)$. It was verified that 
the corresponding variations of \Bs lifetime result in negligible changes in the efficiency.

The validation of Monte Carlo samples is performed by comparing distributions of 
variables used in the event selection between simulation and background-subtracted data.
No significant deviation is found, and thus no systematic uncertainties in the 
efficiency ratio are assigned related to data-simulation discrepancies in those variables. 

\section{Systematic uncertainties}
\label{sec:syst}

Many systematic uncertainties, related to the efficiency of the trigger as well as the reconstruction 
and identification of the muons, cancel out in the measured ratios \rs and \rpp. Since the \bsk and 
\bdk decays have the same number of tracks in the final state, uncertainties related to the track 
reconstruction are of the same size and correlated, and therefore cancel out when propagated to the measured 
ratio \rs. For the ratio \rpp, we consider an additional uncertainty of 4.2\% from the 
uncertainty in the tracking efficiency of two additional pions~\cite{tracking}.

The systematic uncertainty related to the choice of the fit model is evaluated 
by testing different models. The largest deviation in the 
measured ratio from its baseline value is taken as a systematic uncertainty, separately for the variations of the signal and background models.
Several alternative signal models were considered. One is a double Gaussian function for \Bd and \Bs signals with the 
resolution shape fixed to the expectations taken from simulation with only the resolution scaling parameter 
being free in the fit. 
Another signal model is a Student's \textit{t}-distribution~\cite{Studentt} with the 
value of the $n$ parameter fixed to the one measured in simulation.
Alternative background models include polynomials of the second and third degrees, 
an exponential multiplied by a polynomial, and a power function multiplied by an exponential, where 
in all cases the background shape parameters are free to vary in the fits.

\begin{figure}[h!t]
\centering
    \includegraphics[width=0.49\textwidth]{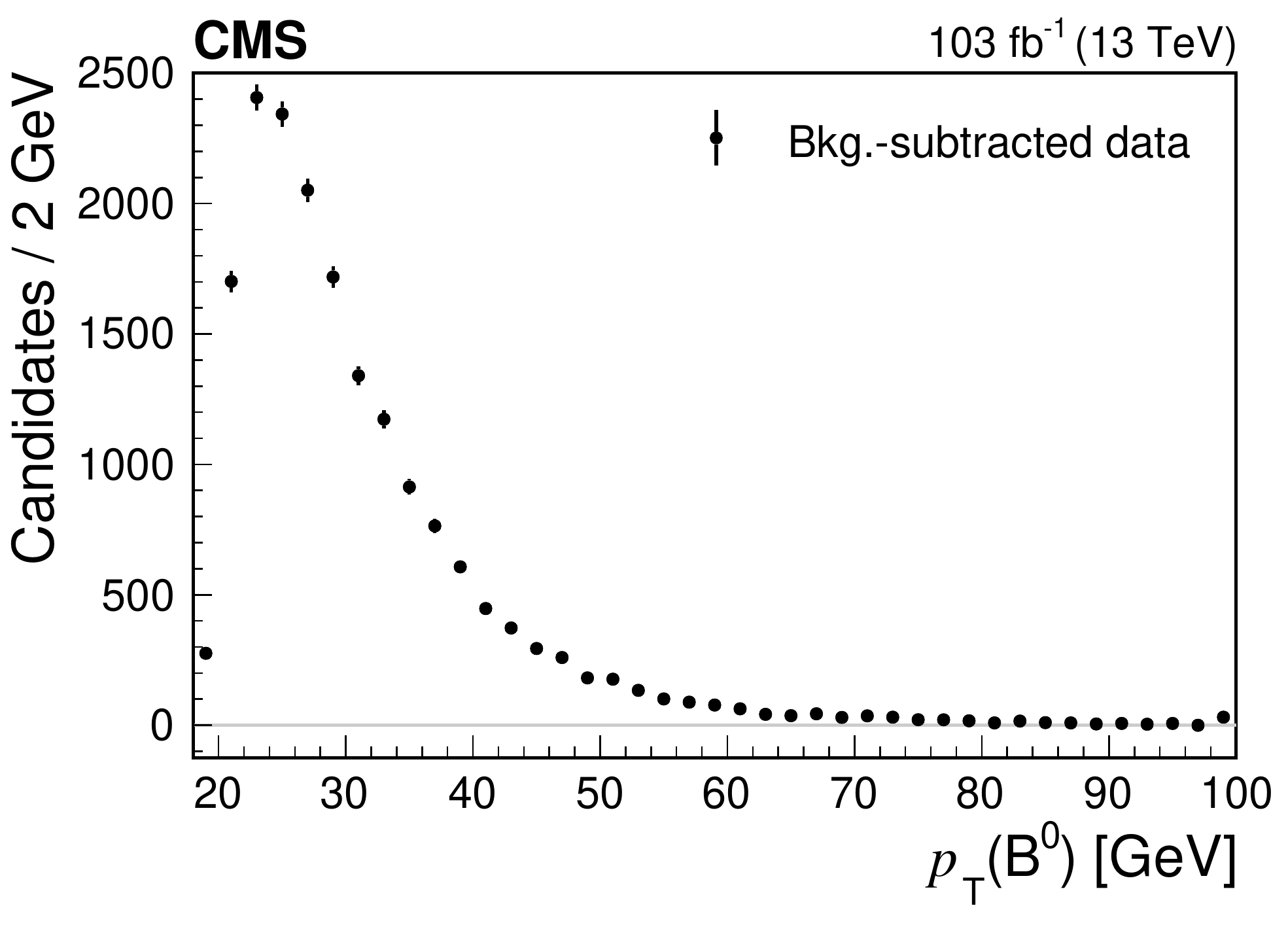}
    \caption{Background-subtracted $\pt(\Bd)$ distribution in data for the \bdk signal. The last bin includes the overflow.
\label{fig:ptB}}
\end{figure}

The uncertainty related to the finite size of the simulation samples (used to measure the 
efficiencies in Section~\ref{sec:eff}) is also considered as a systematic uncertainty.

{\tolerance=2000 
The uncertainty associated with the shape of the \BsPsiKSKPi contribution to the $\psit\KS\pipi$ 
invariant mass distribution is estimated by varying the shape parameters within their uncertainties. 
The largest deviation of \Nbdkpp from the baseline value is 0.5\% which is taken as 
a systematic uncertainty.
\par}

As discussed in Section~\ref{sec5}, the simulation for the \bdkpp decay does 
not take into account the intermediate resonance structure, leading to a significant 
disagreement between data and simulation in the 2- and 3-body mass distributions.
This results in a potential bias in the efficiency reported in Section~\ref{sec:eff}.
To estimate the corresponding systematic uncertainty, the simulated sample is reweighted to be consistent with 
the data, and the difference between the baseline efficiency and the efficiency 
obtained on the weighted sample is taken as a systematic uncertainty.
Due to the limited number of events, it is impossible to assign weights taking 
the ratio of data to simulation in bins of multi-dimensional phase space of the considered 4-body decay.
An iterative procedure has been developed that operates with one-dimensional weights corresponding 
to each 2- and 3-body invariant mass, gradually making the mass distributions on weighted 
simulation sample closer and closer to data, until a satisfactory agreement in all 
intermediate invariant mass distributions is achieved. The distributions of invariant 
masses obtained with the weighted simulation sample are presented in Figs.~\ref{fig:Rewhists2} and~\ref{fig:Rewhists3}.
The efficiency obtained on the weighted simulation sample deviates from the baseline value 
by 5\%, which is taken as a systematic uncertainty due to the intermediate resonance structure.
This efficiency correction procedure with iterative reweighting is verified using a 
dedicated simulation sample instead of data, in which the contributions from \Kstar and \Rho resonances 
are included with arbitrary magnitudes.

All uncertainties described above, excluding the one related to \fsfd for the ratio \rs, 
are summarized in Table~\ref{tab:syst} together with a total systematic uncertainty, 
calculated as a sum in quadrature of the individual sources.

\begin{table}[hb]
    \centering
    \topcaption{Systematic uncertainties (in \%) of the measured branching fraction ratios.}
    \label{tab:syst}
    \begin{tabular}[c]{lcc}
        Source                             &  \rs      &  \rpp     \\[2mm]\hline
        Background model                   &  2.5      &  0.8      \\ 
        Signal model                       &  1.5      &  0.8      \\
        Shape of \BsPsiKSKPi contribution  &  \NA      &  0.5      \\
        Finite size of simulation samples  &  1.3      &  1.1      \\
        Intermediate resonances            &  \NA      &  5.0      \\
        Tracking efficiency                &  \NA      &  4.2      \\[2mm] 
        Total                              &  3.2      &  6.7      \\ 
    \end{tabular}
\end{table}

A measurement of the ratio of the \Bs and \Bd fragmentation fractions, \fsfd, in proton-proton 
collisions at the LHC has been recently reported by the LHCb 
Collaboration~\cite{LHCb:2021qbv}: $\fsfd=(0.263\pm0.008) - (17.6\pm2.1)\times10^{-4}\,\ptb$,
where \ptb in \gev is the transverse momentum of a \B meson produced in 13\TeV proton-proton collisions.
The ratio was found to be independent of the rapidity of the \B meson, but with a significant 
dependence on the transverse momentum of the \B candidate.
The \ptb distribution used in this analysis is shown in Fig.~\ref{fig:ptB}, where the background is 
subtracted using \splot. 
Using the LHCb result and the average \ptb in our events of 31.2\gev, the \fsfd value for 
the kinematic range of this analysis is obtained to be 
$\fsfd = 0.208 \pm 0.007$.
The LHCb \fsfd measurement is mostly dependent on the events with $\pt<20\GeV$, 
while the majority of the events in this analysis have $\pt(\B)>20\GeV$. 
Therefore, we assign an additional systematic uncertainty on \fsfd as the difference 
between 0.208 and the value obtained under the assumption that 
\fsfd becomes constant (0.2278) in the region $\pt>20\gev$. 
This additional uncertainty is estimated to be 0.020, and the total uncertainty on \fsfd
is obtained by summing it in quadrature with the uncertainty of 0.007 obtained above. 
The resulting fragmentation fraction ratio used in the \rs measurement is $\fsfd=0.208\pm0.021$, 
with a relative uncertainty of 10\%.

\section{Measured branching fractions\label{results}}

The branching fraction ratio of the \bsk decay relative to the \bdk 
one is measured using Eq.~\eqref{eq:eqbr} to be
\begin{linenomath}
\ifthenelse{\boolean{cms@external}}
{ % Begin journal version
    \begin{equation*}
    \begin{aligned}
        \rs =& \frac{\bran(\bsk)}{\bran(\bdk)} \\
            =&(3.33 \pm 0.69 \stat \pm 0.11\syst \\
            &\pm 0.34\,(\fsfd)) \times 10^{-2},
    \end{aligned}
    \end{equation*}
} % End journal version
{ % Begin arXiv version
    \begin{equation*}
        \rs = \frac{\bran(\bsk)}{\bran(\bdk)} = (3.33 \pm 0.69 \stat \pm 0.11\syst \pm 0.34\,(\fsfd)) \times 10^{-2},
    \end{equation*}
} % End arXiv version
\end{linenomath}
where the last uncertainty is related to the used value $\fsfd=0.208\pm0.021$. 
Since the knowledge of \fsfd at large \pt can be updated with future measurements, 
allowing to improve the \rs evaluation,
we also provide the measurement of the product 
\begin{linenomath}
\ifthenelse{\boolean{cms@external}}
{ % Begin journal version
    \begin{equation*}
    \begin{aligned}
        \rs\,\frfsfd    &= \frfsfd\,\frac{\bran(\bsk)}{\bran(\bdk)}  \\
                        &= (0.69 \pm 0.14\stat \pm 0.02\syst) \times 10^{-2}.
    \end{aligned}
    \end{equation*}
} % End journal version
{ % Begin arXiv version
    \begin{equation*}
        \rs\,\frfsfd=\frfsfd\,\frac{\bran(\bsk)}{\bran(\bdk)} =
        (0.69 \pm 0.14\stat \pm 0.02\syst) \times 10^{-2}.
    \end{equation*}

} % End arXiv version
\end{linenomath}
In addition, the transverse momentum distribution of the measured \B candidates is presented in Fig.~\ref{fig:ptB} 
and in the HEPData record for this analysis~\cite{hepdata}.

{\tolerance=5200 
The branching fraction ratio of the \bdkpp decay with respect to the \bdk one is measured to be 
\begin{linenomath}
\ifthenelse{\boolean{cms@external}}
{ % Begin journal version
    \begin{equation*}
    \begin{aligned}
      \rpp  &= \frac{\bran(\bdkpp)}{\bran(\bdk)}  \\
            &= 0.480 \pm 0.013\stat \pm 0.032\syst.
    \end{aligned}
    \end{equation*}
} % End journal version
{ % Begin arXiv version
    \begin{equation*}
        \rpp=\frac{\bran(\bdkpp)}{\bran(\bdk)} =
        0.480 \pm 0.013\stat \pm 0.032\syst.
    \end{equation*}

} % End arXiv version
\end{linenomath}
\par}
This ratio is very close to the similar ratio measured with \jpsi instead of \psit~\cite{LHCb:2014sli}.
 
Using the world average value $\bran(\bdk) = (2.90\pm0.25)\times10^{-4}$~\cite{PDG},
the branching fractions of the two newly observed decays are evaluated: 
\begin{linenomath}
\ifthenelse{\boolean{cms@external}}
{ % Begin journal version
    \begin{gather*}
     \bran(\bsk)     \\
     \qquad = (0.97 \pm0.20\stat\pm0.03\syst\\
     \qquad\mathrel{\phantom{=}}\pm0.22\,(\fsfd)\pm0.08\,(\bran))\times 10^{-5}, \\
     \bran(\bdkpp)    \\
     \qquad= (13.9\pm0.4\stat\pm0.9\syst\\
     \qquad\mathrel{\phantom{=}}\pm1.2\,(\bran))\times 10^{-5}, \\
    \end{gather*}
} % End journal version
{ % Begin arXiv version
    \begin{equation*}
    \begin{aligned}
     \bran(\bsk)    &= (0.97 \pm0.20\stat\pm0.03\syst\pm0.22\,(\fsfd)\pm0.08\,(\bran))\times 10^{-5}, \\
     \bran(\bdkpp)  &= (13.9\pm0.4\stat\pm0.9\syst\pm1.2\,(\bran))\times 10^{-5}, \\
    \end{aligned}
    \end{equation*}
} % End arXiv version
\end{linenomath}
where the last uncertainties are from the uncertainty in $\bran(\bdk)$.

\section{Summary\label{summary}}
The \bsk and \bdkpp decays are observed using proton-proton collision data collected by the CMS 
experiment at 13\TeV with an integrated luminosity of 103\fbinv.
Their branching fractions are measured with respect to the \bdk decay to be 
$\bran(\bsk)/\bran(\bdk) = (3.33 \pm 0.69 \stat \pm 0.11\syst \pm 0.34\,(\fsfd)) \times 10^{-2}$, 
and $\bran(\bdkpp)/\bran(\bdk)=0.480 \pm 0.013\stat \pm 0.032\syst$, 
where the last uncertainty in the first ratio corresponds to the uncertainty in 
the ratio of production cross sections of \Bs and \Bd mesons.
The 2- and 3-body invariant mass distributions of the \bdkpp decay products
do not show significant exotic narrow structures in addition to the known 
light meson resonances.
Further studies with more data will be needed to investigate more precisely 
the internal dynamics of the \bdkpp decay, and to perform \textit{CP} asymmetry measurements 
in the two observed decays in the future.

\begin{acknowledgments}
    We congratulate our colleagues in the CERN accelerator departments for the excellent performance of the LHC and thank the technical and administrative staffs at CERN and at other CMS institutes for their contributions to the success of the CMS effort. In addition, we gratefully acknowledge the computing centres and personnel of the Worldwide LHC Computing Grid and other centres for delivering so effectively the computing infrastructure essential to our analyses. Finally, we acknowledge the enduring support for the construction and operation of the LHC, the CMS detector, and the supporting computing infrastructure provided by the following funding agencies: BMBWF and FWF (Austria); FNRS and FWO (Belgium); CNPq, CAPES, FAPERJ, FAPERGS, and FAPESP (Brazil); MES and BNSF (Bulgaria); CERN; CAS, MoST, and NSFC (China); MINCIENCIAS (Colombia); MSES and CSF (Croatia); RIF (Cyprus); SENESCYT (Ecuador); MoER, ERC PUT and ERDF (Estonia); Academy of Finland, MEC, and HIP (Finland); CEA and CNRS/IN2P3 (France); BMBF, DFG, and HGF (Germany); GSRI (Greece); NKFIA (Hungary); DAE and DST (India); IPM (Iran); SFI (Ireland); INFN (Italy); MSIP and NRF (Republic of Korea); MES (Latvia); LAS (Lithuania); MOE and UM (Malaysia); BUAP, CINVESTAV, CONACYT, LNS, SEP, and UASLP-FAI (Mexico); MOS (Montenegro); MBIE (New Zealand); PAEC (Pakistan); MSHE and NSC (Poland); FCT (Portugal); JINR (Dubna); MON, RosAtom, RAS, RFBR, and NRC KI (Russia); MESTD (Serbia); MCIN/AEI and PCTI (Spain); MOSTR (Sri Lanka); Swiss Funding Agencies (Switzerland); MST (Taipei); ThEPCenter, IPST, STAR, and NSTDA (Thailand); TUBITAK and TAEK (Turkey); NASU (Ukraine); STFC (United Kingdom); DOE and NSF (USA).

    \hyphenation{Rachada-pisek} Individuals have received support from the Marie-Curie programme and the European Research Council and Horizon 2020 Grant, contract Nos.\ 675440, 724704, 752730, 758316, 765710, 824093, 884104, and COST Action CA16108 (European Union); the Leventis Foundation; the Alfred P.\ Sloan Foundation; the Alexander von Humboldt Foundation; the Belgian Federal Science Policy Office; the Fonds pour la Formation \`a la Recherche dans l'Industrie et dans l'Agriculture (FRIA-Belgium); the Agentschap voor Innovatie door Wetenschap en Technologie (IWT-Belgium); the F.R.S.-FNRS and FWO (Belgium) under the ``Excellence of Science -- EOS" -- be.h project n.\ 30820817; the Beijing Municipal Science \& Technology Commission, No. Z191100007219010; the Ministry of Education, Youth and Sports (MEYS) of the Czech Republic; the Deutsche Forschungsgemeinschaft (DFG), under Germany's Excellence Strategy -- EXC 2121 ``Quantum Universe" -- 390833306, and under project number 400140256 - GRK2497; the Lend\"ulet (``Momentum") Programme and the J\'anos Bolyai Research Scholarship of the Hungarian Academy of Sciences, the New National Excellence Program \'UNKP, the NKFIA research grants 123842, 123959, 124845, 124850, 125105, 128713, 128786, and 129058 (Hungary); the Council of Science and Industrial Research, India; the Latvian Council of Science; the Ministry of Science and Higher Education and the National Science Center, contracts Opus 2014/15/B/ST2/03998 and 2015/19/B/ST2/02861 (Poland); the Funda\c{c}\~ao para a Ci\^encia e a Tecnologia, grant CEECIND/01334/2018 (Portugal); the National Priorities Research Program by Qatar National Research Fund; the Ministry of Science and Higher Education, projects no. 0723-2020-0041 and no. FSWW-2020-0008, and the Russian Foundation for Basic Research, project No.19-42-703014 (Russia); MCIN/AEI/10.13039/501100011033, ERDF ``a way of making Europe", and the Programa Estatal de Fomento de la Investigaci{\'o}n Cient{\'i}fica y T{\'e}cnica de Excelencia Mar\'{\i}a de Maeztu, grant MDM-2017-0765 and Programa Severo Ochoa del Principado de Asturias (Spain); the Stavros Niarchos Foundation (Greece); the Rachadapisek Sompot Fund for Postdoctoral Fellowship, Chulalongkorn University and the Chulalongkorn Academic into Its 2nd Century Project Advancement Project (Thailand); the Kavli Foundation; the Nvidia Corporation; the SuperMicro Corporation; the Welch Foundation, contract C-1845; and the Weston Havens Foundation (USA).
\end{acknowledgments}

\bibliography{auto_generated}
\cleardoublepage \appendix\section{The CMS Collaboration \label{app:collab}}\begin{sloppypar}\hyphenpenalty=5000\widowpenalty=500\clubpenalty=5000\cmsinstitute{Yerevan~Physics~Institute, Yerevan, Armenia}
A.~Tumasyan
\cmsinstitute{Institut~f\"{u}r~Hochenergiephysik, Vienna, Austria}
W.~Adam\cmsorcid{0000-0001-9099-4341}, J.W.~Andrejkovic, T.~Bergauer\cmsorcid{0000-0002-5786-0293}, S.~Chatterjee\cmsorcid{0000-0003-2660-0349}, K.~Damanakis, M.~Dragicevic\cmsorcid{0000-0003-1967-6783}, A.~Escalante~Del~Valle\cmsorcid{0000-0002-9702-6359}, R.~Fr\"{u}hwirth\cmsAuthorMark{1}, M.~Jeitler\cmsAuthorMark{1}\cmsorcid{0000-0002-5141-9560}, N.~Krammer, L.~Lechner\cmsorcid{0000-0002-3065-1141}, D.~Liko, I.~Mikulec, P.~Paulitsch, F.M.~Pitters, J.~Schieck\cmsAuthorMark{1}\cmsorcid{0000-0002-1058-8093}, R.~Sch\"{o}fbeck\cmsorcid{0000-0002-2332-8784}, D.~Schwarz, S.~Templ\cmsorcid{0000-0003-3137-5692}, W.~Waltenberger\cmsorcid{0000-0002-6215-7228}, C.-E.~Wulz\cmsAuthorMark{1}\cmsorcid{0000-0001-9226-5812}
% ...
\vskip\cmsinstskip
% ...
1:~Also at TU Wien, Wien, Austria\\
% ...
\end{sloppypar}
%%% END EDITABLE REGION %%%
% skeleton_end
\end{document}